# Exploring the Horizon: A Comprehensive Survey of Rowhammer

Amir Naseredini[1,2] [0009-0009-3836-3361]

[1] University of Sussex, Brighton, UK
[2] Canonical Group Ltd, London, UK

**Abstract.** *Rowhammer* poses a significant security challenge for modern computers, specifically affecting Dynamic Random Access Memory (DRAM). Given society's growing reliance on computer systems, ensuring the reliability of hardware is of utmost importance. This paper provides a comprehensive survey of *Rowhammer*, examining the literature from various angles. We categorise studies on *Rowhammer* into attacks, defences, and intriguing work, exploring each category in detail. Furthermore, we classify papers within each category into distinct yet overlapping classes and present an overview of the papers in each class.

## 1 Introduction

*Rowhammer*, introduced in 2014 [35], is a security attack that exploits the physical characteristics of DRAM. DRAMs in modern computers use a single capacitor per bit, which can be affected by electromagnetic interference caused by accessing nearby bits. This can lead to bit flips in adjacent cells. Factors such as data stored in the neighbourhood and frequency of accesses contribute to the probability of bit flips. Refreshing the memory cells reduces the probability of bit flips temporarily. Refer to [20] and [22] for details on executing a *Rowhammer* attack.

The study of *Rowhammer* began with the work of Kim et al. in [35] in 2014. Since then, concerns have been raised regarding the security of systems where the DRAM, a critical and vulnerable component, is susceptible to *Rowhammer* attacks. This issue has received significant attention, leading to numerous projects and research papers investigating various attack and defence techniques.

This work focuses on discussing *Rowhammer* Attack Variants in Section 2, *Rowhammer* Defence Variants in Section 3, and presents additional noteworthy research on *Rowhammer* from different perspectives in Section 4.

## 2 Attack Variants

The initial in-depth discussions and analyses of *Rowhammer* were initiated by Kim et al. in their work [35]. They conducted a study on the behaviour of DRAM disturbance errors using a platform based on Field-Programmable Gate Array (FPGA). In addition to their findings, they proposed seven potential mitigations, including Probabilistic Adjacent Row Activation (PARA). PARA is a mitigation technique with low overhead that aims to prevent *Rowhammer* by probabilistically issuing refreshes to rows considered potential victims. It is designed to have minimal impact on performance but has certain limitations, such as the requirement for address mapping between logical and physical addresses in DRAM, which is considered proprietary information for manufacturers. The paper also introduces a user-level program that exploits the vulnerability. They explain that by reading rows from DRAM, a program may be able to bypass the promised memory protection and compromise the integrity of data stored in DRAM. Furthermore, they provide insights into the main cause of disturbance errors, which ultimately lead to the occurrence of *Rowhammer*.

In the upcoming subsections, we will examine the significant papers published in each respective year. Following that, we will introduce additional categories for the same list of papers as we conclude this section. It is important to note that these categories are not mutually exclusive.

### 2.1 2015

Shortly after [35], in 2015, Seaborn and Dullien introduced a new *Rowhammer* attack, successfully gaining kernel privileges from an unprivileged user-space program [64]. This work revealed the widespread presence of exploitable disturbance errors in commodity DRAM. The attack involved corrupting Page Table Entry (PTE)s using *Rowhammer*, granting the attacker complete read and write access to the entire memory.

### 2.2 2016

Two years after *Rowhammer*'s introduction, interesting papers emerged. One such paper, in [5], presented the first JavaScript-based *Rowhammer* attack. It exploited the vulnerability remotely within Microsoft Edge browser, utilizing memory deduplication. In the same year, *Rowhammer.js* was introduced [22]. It employed cache eviction techniques to flush accessed rows from the cache and explored cache eviction parameters. The work featured *Rowhammer* implementations in both JavaScript and native code.

In the same year, *Flip Feng Shui* [57] was introduced as a powerful attack vector capable of controlled exploitation. It relied on flipping bits in memory pages and reverse-mapping physical to virtual addresses. The authors applied *Flip Feng Shui* to Virtual Machine (VM)s, breaking their memory isolation. Another instance, *Drammer* [70], targeted Android devices using *Flip Feng Shui*. Authors studied *Flip Feng Shui* and found three conditions necessary for a successful *Flip Feng Shui Rowhammer* attack: rapid hammering, placing victim's data in a vulnerable memory section (a.k.a., *memory massaging*), and targeting contiguous physical memory addresses. *Drammer* became the first Android

*Rowhammer* attack, capable of being perform by any app without special permissions and independent of software-level vulnerabilities in Android.

In another work, [77], a combination of *Rowhammer* and virtualisation was demonstrated, focusing on enabling a VM to have read and write access to the entire memory. They implemented a double-sided *Rowhammer* attack between guest VMs, showcasing information leakage. This work is partially related to [57] in terms of performing a *Rowhammer* attack to break memory isolation. Qiao and Seaborn in [54] introduced a *Rowhammer* attack without using the *CLFLUSH* instruction, enabling a new remote *Rowhammer*. They discovered that commonly used functions like *memset* and *memcpy* could trigger *Rowhammer*, bypassing sandboxes and existing defences in the application layer. Another paper [4] used eviction sets to perform the *Rowhammer* attack, similar to [22]. And, *DRAMA* [51] presented two approaches to aid reverse engineering of DRAM address mappings, targeting VMs. Lastly, *Rowhammer: Reliability analysis and security implications* [36], summarises the idea of the original *Rowhammer* paper in [35].

### 2.3 2017

In the next paper [27], the drop-and-lock policy of the Central Processing Unit (CPU) is exploited through an unprivileged *Rowhammer* attack to halt the CPU state and conduct a Denial of Service (DoS) attack. Aga et al. [1] focused on Cache Allocation Technology (CAT) as a defence mechanism against DoS attacks in inter-VM scenarios. And, they demonstrate the first single-sided and *CLFLUSH*-free *Rowhammer* attack. Their attack bypasses several protection mechanisms, such as doubling the DRAM refresh rate, restricting access to virtual memory mapping, and limiting cache flush operations. This attack targets VMs on the same host machine. Another study in [14] aimed to survey attacks, including *Rowhammer*, on real-time embedded systems, providing a comprehensive overview of various *Rowhammer* attacks up to the publication date.

### 2.4 2018

In [16], authors demonstrate the first *Rowhammer* attack executed from the Graphics Processing Unit (GPU) to target a mobile phone's browser. This reliable, end-to-end remote attack also explored the potential of integrated GPUs for microarchitectural attacks. In the next paper, [21], two significant contributions were made. Firstly, they introduced a new variant of *Rowhammer* called one-location hammering, where a bit flip in a row could be induced by hammering only one row in DRAM. Secondly, the authors proposed a new categorization for *Rowhammer* defences, classifying them into five categories: Static Analysis, Monitoring Performance Counters, Monitoring Memory Access Patterns, Preventing Exhaustion-based Page Placement, and Preventing Physical Proximity to Kernel Pages.

The next two studies, Throwhammer [68] and Nethammer [42], aimed to perform *Rowhammer* attacks over a network. Throwhammer evaluates hosts remotely without code execution permission, showcasing the first remote *Rowhammer* attack achieved solely through network packets. Nethammer demonstrates the ineffectiveness of Target Row Refresh (TRR) techniques and introduced the first one-location *Rowhammer* attack variant on *ARM* machines.

Another VM attack was presented in [52], featuring deterministic single-sided and double-sided *Rowhammer* attacks to compromise private keys, with a focus on cryptographic primitives. Tatar et al. bypass ANVIL [2] and CATT [6] in [67] and demonstrate highly effective *Rowhammer* attacks, discussing physical and DRAM address spaces. The study in [71] showcased *RAMpage*, a series of *Rowhammer* attacks on Android devices, including root exploits and app-to-app attacks that bypassed mitigation mechanisms. And [7] examined the reliability of OpenSSL against *Rowhammer*-induced faults, successfully restoring Rivest–Shamir–Adleman (RSA) private keys through two analyses. The last paper of 2018, [82], focuses on *ARMv8-A* cache and *Rowhammer* attacks. The authors revealed an unknown CPU instruction usable to exploit *Rowhammer* and evaluated its impacts.

### 2.5 2019

In 2019, [11] challenged the effectiveness of Error Correction Code (ECC) as a complete mitigation against *Rowhammer*. They reverse-engineered ECC implementations, exposing vulnerabilities, and introduced *ECCploit*, a novel *Rowhammer* attack that bypasses ECC mitigation. Zhang et al. presented TeleHammer in [83], a new family of *Rowhammer* attacks. They addressed the need for direct access to the hammer row(s) in existing attacks and introduced *TeleHammer*, which triggers the hammer row(s) through a third party. They showcased *PThammer*, an instance of *TeleHammer*, exploiting *Rowhammer* using the CPU's address-translation component.

### 2.6 2020

In [9], the authors present a testing approach for evaluating DRAM devices against *Rowhammer*, with a focus on cloud providers. They address the challenges of finding an optimal instruction sequence to activate memory rows and identifying the physical neighbourhood relations between rows. The paper discusses successful *Rowhammer* attacks on TRR-DRAMs and questions the effectiveness of TRR or its potential lack of enablement. Detailed technical information is provided.

*RAMBleed* [39] demonstrates the first *Rowhammer* attack targeting DRAM confidentiality. It challenges the assumption that *flipping bits in own memory pages wouldn't violate security.* They recover a secret key from an OpenSSH server as a case study. Next, [81] presents *DeepHammer*, the first *Rowhammer* attack against Quantised Deep Neural Network (QDNN), compromising it entirely. Two main challenges are considered: optimizing marked bits and successfully attacking the target bits. This attack is the first to exploit a hardware fault to compromise Deep Neural Network (DNN) deterministically.

*TRRespass* [17] thoroughly investigates TRR and demonstrates *Rowhammer* exploits on various DRAM models, including mobile devices. Their tool, *TRRespass*, bypasses TRR policies using many-sided *Rowhammer* attacks. In [31], experiments reveal that *Rowhammer* is worsening over time, with newer versions requiring fewer accesses to an aggressor row. The paper emphasises the need for more effective mitigations by studying across a wide range of DRAM chips. GhostKnight [84] is the first to combine *Spectre* and *Rowhammer*, compromising the victim and violating integrity. It introduces the *GhostKnight* attack, using *Spectre* to access hammer rows.

### 2.7 2021

*SMASH* [58] builds upon *TRRespass* [17] by using many-sided *Rowhammer* for a remote attack from JavaScript on TRR-protected memories. The paper addresses challenges of memory allocation, generating access patterns, and timing. It also introduces a mechanism for generating *Rowhammer* access patterns without cache flush. In [53], a new attack vector for *Rowhammer* is introduced, demonstrating the impact of far neighbours on a victim row. The final paper of the section, *DeepSteal* [56], presents *HammerLeak* for parameter leaking from DNNs. The end-to-end framework, *DeepSteal*, employs new techniques and algorithms. HammerLeak shares similarities with *RAMBleed* [39].

### 2.8 2022

In 2022, *Blacksmith* fuzzer was introduced in [28]. It generates access patterns, including non-uniform ones, for *Rowhammer* attacks. The authors evaluated *Blacksmith* against different memory types. *SpecHammer* [69] combines *Spectre* and *Rowhammer*, utilizing speculative execution to enhance the attack. They achieved higher bit flip rates and discovered numerous gadgets in the Linux kernel. SpecHammer gadgets are used for serious attacks such as finding canaries for buffer overflow attacks and accessing any part of the memory. *Half-Double* [37], a sequel to [53], combines *Spectre* with *Rowhammer* as well. It leverages TRR to compromise up-to-date *Rowhammer* mitigations. The authors evaluate susceptible devices and modules.

Authors of [44] demonstrate that a *Rowhammer* attack can occur without using malicious looking code. They identify this behaviour in Intel *ccNUMA* protocols and propose a new protocol, *MOESI-prime*, to address the issue. *HammerScope* [8] is a *Rowhammer* attack that leaks kernel memory and bypasses Kernel Address Space Layout Randomization (KASLR). They also show that there are correlations between vulnerable bits power consumption patterns. *PROTRR* [45] introduces the *FEINTING* attack and proves its optimality against ideal TRR mitigation. And *Juggernaut* [75] breaks the Randomized Row-Swap (RRS) mitigation from [59]. *Implicit Hammer* [85] eliminates the need for direct access to a hammer row for Rowhammer attacks. They introduce two new types of implicit hammer attacks: *PThammer* and *SyscallHammer*.

### 2.9 Attacks in Categories

In summarizing the review section on *Rowhammer* attacks, we categorise the relevant works into non-mutually exclusive categories: 1) Open Source Code, 2) JavaScript-based, 3) Exploiting Memory Deduplication, 4) Within acVMs, 5) Utilizing FPGA, 6) Cache Eviction Techniques, 7) DoS, 8) Impacting Confidentiality, 9) Targeting Mobile Phones, 10) Remote Attacks, and 11) *Rowhammer* combined with *Spectre*.

***Rowhammer* Attacks with Open Source Code.** Several works have released their implementation publicly. This can be important for the community to verify the work and follow up on it. Table 1 illustrates the list of papers with open source implementations.

| | Paper(s) |
|---|---|
| 1. | Exploiting the DRAM rowhammer bug to gain kernel privileges [64] |
| 2. | Rowhammer. js: A remote software-induced fault attack in javascript [22] |
| 3. | Drammer: Deterministic rowhammer attacks on mobile platforms [70] |
| 4. | Throwhammer: Rowhammer attacks over the network and defenses [68] |
| 5. | Defeating software mitigations against rowhammer: a surgical precision hammer [67] |
| 6. | Half-Double: Hammering From the Next Row Over [37] |
| 7. | MOESI-prime: preventing coherence-induced hammering in commodity workloads [44] |
| 8. | HammerScope: Observing DRAM Power Consumption Using Rowhammer [8] |

**Table 1.** *Rowhammer* Attacks with Open Source Code

***Rowhammer* Attacks in JavaScript.** There are multiple attacks in different programming languages, however, JavaScript is an interesting choice of programming language, particularly because of its nature. List of papers introduced an attack in JavaScript is presented in Table 2.

| | Paper(s) |
|---|---|
| 1. | Dedup est machina: Memory deduplication as an advanced exploitation vector [5] |
| 2. | Rowhammer. js: A remote software-induced fault attack in javascript [22] |
| 3. | Grand pwning unit: Accelerating microarchitectural attacks with the GPU [16] |
| 4. | {SMASH}: Synchronized Many-sided Rowhammer Attacks from JavaScript [58] |

**Table 2.** *Rowhammer* Attacks in JavaScript

***Rowhammer* Attacks using Memory Deduplication.** Memory deduplication is a technique introduced to use the memory in a more efficient way. However, everything comes with a price and there are *Rowhammer* attacks utilising memory deduplication. Table 3 below shows the lists of such attacks.

| | Paper(s) |
|---|---|
| 1. | Flip Feng Shui: Hammering a needle in the software stack [57] |
| 2. | Dedup est machina: Memory deduplication as an advanced exploitation vector [5] |
| 3. | Drammer: Deterministic rowhammer attacks on mobile platforms [70] |

**Table 3.** *Rowhammer* Attacks using Memory Deduplication

***Rowhammer* Attacks in VMs.** Papers introduced in Table 4 demonstrate *Rowhammer* attacks on VMs. Also, the attack introduced in [4] is not demonstrated on VM, however, authors claim that it is possible to run it in VM.

| | Paper(s) |
|---|---|
| 1. | Flip Feng Shui: Hammering a needle in the software stack [57] |
| 2. | One bit flips, one cloud flops: Cross-vm row hammer attacks and privilege escalation [77] |
| 3. | DRAMA: Exploiting DRAM Addressing for Cross-CPU Attacks [51] |
| 4. | Curious case of rowhammer: flipping secret exponent bits using timing analysis [4] |
| 5. | Attacking deterministic signature schemes using fault attacks [52] |

**Table 4.** *Rowhammer* Attacks in Virtual Machine (VM)

***Rowhammer* Attacks using FPGA.** Papers using actual hardware either use the real-world environments or FPGA boards. One of the advantages of using FPGA boards is to eliminate unnecessary noise and focus on the main purpose of the attack. Table 5 shows the list of papers using FPGA in order to run their attacks, however, this does not mean that their attacks cannot be done on real-world machines.

| | Paper(s) |
|---|---|
| 1. | Flipping Bits in Memory Without Accessing Them: An Experimental Study of DRAM Disturbance Errors [35] |
| 2. | Exploiting correcting codes: On the effectiveness of ecc memory against rowhammer attacks [11] |
| 3. | Are We Susceptible to Rowhammer? An End-to-End Methodology for Cloud Providers [9] |
| 4. | TRRespass: Exploiting the Many Sides of Target Row Refresh [17] |
| 5. | Revisiting rowhammer: An experimental analysis of modern dram devices and mitigation techniques [31] |
| 6. | Introducing Half-Double: New hammering technique for DRAM Rowhammer bug [53] |
| 7. | Blacksmith: Scalable rowhammering in the frequency domain [28] |
| 8. | Half-Double: Hammering From the Next Row Over [37] |

**Table 5.** *Rowhammer* Attacks using FPGA

***Rowhammer* Attacks using Cache Eviction.** There are several technique in order to remove the accessed row from the cache for the purpose of having another access to the row in the memory. One of these techniques is to use cache eviction policies. Papers introducing attacks using cache eviction techniques are listed in Table 6.

| | Paper(s) |
|---|---|
| 1. | Curious case of rowhammer: flipping secret exponent bits using timing analysis [4] |
| 2. | Rowhammer. js: A remote software-induced fault attack in javascript [22] |
| 3. | Grand pwning unit: Accelerating microarchitectural attacks with the GPU [16] |
| 4. | Triggering Rowhammer Hardware Faults on ARM: A Revisit [82] |
| 5. | {SMASH}: Synchronized Many-sided Rowhammer Attacks from JavaScript [58] |

**Table 6.** *Rowhammer* Attacks using Cache Eviction

***Rowhammer* Attacks with DoS impact.** Each attack is designed to challenge one (or more) security feature(s). One of the targets has always been the availability of the system to the user. *Rowhammer* attacks with the purpose of causing a Denial of Service (DoS), are illustrated in Table 7.

| | Paper(s) |
|---|---|
| 1. | Sgx-bomb: Locking down the processor via rowhammer attack [27] |
| 2. | When good protections go bad: Exploiting anti-DoS measures to accelerate Rowhammer attacks [1] |

**Table 7.** *Rowhammer* Attacks with DoS impact

***Rowhammer* Attacks on Confidentiality.** Another target of the attacks is the confidentiality of the system and there are *Rowhammer* attacks targeting it. These attacks are shown in Table 8.

| | Paper(s) |
|---|---|
| 1. | Kwong, Andrew and Genkin, Daniel and Gruss, Daniel and Yarom, Yuval [39] |
| 2. | DeepSteal: Advanced Model Extractions Leveraging Efficient Weight Stealing in Memories [56] |
| 3. | HammerScope: Observing DRAM Power Consumption Using Rowhammer [8] |

**Table 8.** *Rowhammer* Attacks on Confidentiality

***Rowhammer* Attacks on Mobile Phone.** Table 9 lists the *Rowhammer* attacks that target mobile devices. These attacks are designed for this purpose to exploit mobile device DRAMs.

| | Paper(s) |
|---|---|
| 1. | Drammer: Deterministic rowhammer attacks on mobile platforms [70] |
| 2. | Grand pwning unit: Accelerating microarchitectural attacks with the GPU [16] |
| 3. | Triggering Rowhammer Hardware Faults on ARM: A Revisit [82] |
| 4. | TRRespass: Exploiting the Many Sides of Target Row Refresh [17] |
| 5. | GuardION: Practical mitigation of DMA-based rowhammer attacks on ARM [71] |
| 6. | Introducing Half-Double: New hammering technique for DRAM Rowhammer bug [53] |
| 7. | Blacksmith: Scalable rowhammering in the frequency domain [28] |

**Table 9.** *Rowhammer* Attacks on Mobile Phone

**Remote *Rowhammer* Attacks.** *Rowhammer* can be done both locally and remotely. There are several number of *Rowhammer* attacks in which they can exploit the target machine remotely. These attacks are listed in Table 10.

| | Paper(s) |
|---|---|
| 1. | Dedup est machina: Memory deduplication as an advanced exploitation vector [5] |
| 2. | Rowhammer. js: A remote software-induced fault attack in javascript [22] |
| 3. | A new approach for rowhammer attacks [54] |
| 4. | Sgx-bomb: Locking down the processor via rowhammer attack [27] |
| 5. | Grand pwning unit: Accelerating microarchitectural attacks with the GPU [16] |
| 6. | Throwhammer: Rowhammer attacks over the network and defenses [68] |
| 7. | Nethammer: Inducing rowhammer faults through network requests [42] |
| 8. | {SMASH}: Synchronized Many-sided Rowhammer Attacks from JavaScript [58] |
| 9. | HammerScope: Observing DRAM Power Consumption Using Rowhammer [8] |

**Table 10.** Remote *Rowhammer* Attacks

**Attacks with *Rowhammer* + Spectre.** Several attacks were introduced as a combination of *Rowhammer* attack and Spectre attack. The essence of this type of attacks is to leverage the power of both attacks in order to introduce a more powerful one. The list of attacks in this category is shown in Table 11.

| | Paper(s) |
|---|---|
| 1. | Ghostknight: Breaching data integrity via speculative execution [84] |
| 2. | Spechammer: Combining spectre and rowhammer for new speculative attacks [69] |
| 3. | Half-Double: Hammering From the Next Row Over [37] |
| 4. | HammerScope: Observing DRAM Power Consumption Using Rowhammer [8] |

**Table 11.** Attacks with *Rowhammer* + Spectre

## 3 Defence Variants

Kim et al. 's original work on *Rowhammer* [35] also provided suggestions for potential mitigations to address the issue. Here are their proposed ideas:

- Enhancing the design of *DRAM* chips to resolve the problem
- Utilizing technologies like ECC to detect and correct unexpected bit flips
- Reducing the refresh interval of DRAM to refresh frequently accessed bits and prevent charge loss caused by *Rowhammer*
- Manufacturers can conduct tests to identify potential victim rows and avoid using them by mapping them to spare cells

– DRAM users can run tests to identify vulnerable cells and apply appropriate mitigations
– Identifying frequently accessed rows and issuing refreshes to their potential victim rows
– Implementing PARA, their primary suggested technique for mitigating Rowhammer. PARA involves probabilistically refreshing a neighbouring row each time any row is accessed. With PARA in place, when an aggressor row is accessed multiple times, there is a high probability that its adjacent rows have been refreshed, effectively mitigating *Rowhammer*

In a parallel publication from 2014 [30], two hardware-based mitigation approaches were proposed as countermeasures to *Rowhammer*, namely Counter-Based Row Activation (CRA) and Probabilistic Row Activation (PRA). These ideas resemble the concepts presented in [35]. CRA suggests monitoring row accesses and, upon detecting a potential bit flip threat from an aggressor row, refreshing the potential victim rows. On the other hand, PRA, similar to PARA proposed in [35], suggests refreshing adjacent rows of any accessed row with a low probability.

We will now proceed to the subsequent subsections, organized chronologically according to the publication years, where we will present the *Rowhammer* literature from each year. Towards the conclusion of this section, we will introduce a set of non-exclusive categories that encompass the related papers.

### 3.1 2016

In [57], mitigation ideas are proposed for *Rowhammer* attacks. Two types of mitigations are suggested: hardware-based changes and software-based changes. Hardware-based mitigations use counters to track accesses and refresh potential victims. Where, software-based mitigations involve disabling memory deduplication and using randomness for physical page allocation. The study in [70] introduces a defence mechanism with three ideas to prevent *Rowhammer* attacks: limiting permissions of user-level applications on mobile devices, implementing physical separation in memory, and implementing stricter measures to prevent attackers from finding vulnerable areas and exhausting memory.

*ANVIL* [2] is a software-based mitigation mechanism that monitors last-level cache misses to detect suspicious *Rowhammer* activities. It refreshes potential victim rows with high cache miss rates. In [65], a hardware-based approach called Counter-Based Tree (CBT) is proposed. It dynamically partitions and tracks potentially aggressive rows. Authors in [6] introduce *B-CATT* and *G-CATT* as separate software-level mitigation mechanisms. *B-CATT* deactivates susceptible memory parts, while *G-CATT* enforces physical separation. The next study in [25] presents MicroArchitectural Side Channel Attack Trapper (MASCAT), a configurable static analyser tool that detects characteristics of target attacks in binary inputs. And in [18], a hardware-based solution suggests adding *dummy cells* to divert the attacker's attention from actual information-containing cells.

### 3.2 2017

*Secure Page Fusion with VUsion* [49] aims to prevent *Rowhammer* by disguising fused and non-fused pages by introducing *VUsion*. In [66], a hardware-based technique called Probabilistic Rowhammer HIstory Table (PRoHIT) is proposed. It tracks accesses to enhance the effectiveness of PARA. Next, [41], introduces Time Window Counter based row refresh (TWiCe), a hardware-based defence that monitors accessed rows to detect potential *Rowhammer* activity. In [26], two separate mitigation techniques are presented. *Air-Gap* focuses on physical-level separation between allocated memories, while *Row-Mix* suggests randomly mixing row addresses within the DRAM module to hinder address mapping.

### 3.3 2018

In *Another flip in the wall of rowhammer defenses* [21], a new *Rowhammer* defence classification is discussed, and categorised are discussed in Attack Variants in Section 2. The work in [68] introduces ALlocations ISolated (ALIS), a tool for safe allocation of Direct Memory Access (DMA) buffers and guard zones to effectively prevent *Rowhammer* attacks with minimal overhead. *GuardION* [71] is an open-source mitigation mechanism that prevents *Rowhammer* attacks using DMA, which are common in mobile devices. *ZebRAM* [38] is an open-source software-level *Rowhammer* mitigation that separates the attacker row from potential victims. Finally, [72] proposes a hardware-based mitigation technique with a sliding window and a dynamic tree to track potential victim rows and detect bit flips.

### 3.4 2019

In [40], an extension of [41] is introduced, optimizing the counter-based technique at the hardware level. Authors in [76] propose Cell-Type-Aware (CTA), a memory allocation technique in the Operating System (OS) level, effectively preventing PTE-based *Rowhammer* attacks. And the next study, [34], presents a hardware-based technique that remaps addresses and uses separate mappings for different DRAMs to mitigate *Rowhammer*. In the hardware-based technique introduced in [74], authors propose Disturbance Bin Counter (DBC) to detect disturbance errors in DRAM by monitoring accesses and refreshes.

### 3.5 2020

The only paper we cover from 2020 is *Graphene* [50], a scalable hardware-based mitigation mechanism against the *Rowhammer* attack. It showcases how this technique effectively monitors memory accesses to identify potential victim rows and proactively mitigates attacks before any bit flips occur.

### 3.6 2021

The authors of *BlockHammer* [79] propose a new classification for *Rowhammer* mitigation techniques: increased refresh rate, physical isolation, reactive refresh, and proactive throttling. *BlockHammer* treats DRAM as a black box and consists of *RowBlocker* and *AttackThrottler* as the main components to stop *Rowhammer* attacks. They also introduce the *Rowhammer* likelihood index. Another work, [43], suggests somehow similar categories: isolation-centric, frequency-centric, and refresh-centric, with introduction to defence mechanisms needed to be designed at both hardware and software levels. *Panopticon* [3] is a hardware-based technique that tracks accesses using counters to detect and prevent bit flips. And *Mithril* [33] is compatible with ReFresh Management (RFM) and uses counters for mitigation. The Silver Bullet Technique [80] defines a threshold and proactively refreshes potential victim rows. *HammerFilter* [32] classifies attacks as counter-based or probabilistic methods and introduces *HammerFilter* which combines probabilistic behaviour with counters, using Counting Bloom Filter (CBF)s.

### 3.7 2022

Authors of [59] introduce RRS as a hardware-based mitigation mechanism for *Rowhammer*. It utilizes Hot-Row Tracker (HRT) and Row-Indirection Table (RIT) to track accesses, identify suspicious activities, and proactively mitigate the attack. *PROTRR* [45] focuses on TRR mitigation and considers RFM compatibility. It is a hardware-based defence mechanism using counters. In [29], a combination of hardware-based and software-based approach using Message Authentication Code (MAC) is proposed to detect and correct bit flips. *Hydra* [55] is a hardware-based defence mechanism that tracks accesses to prevent *Rowhammer* attacks. And Hidden Row Activation (HiRA) [78] reduces refresh latency and increases the RowHammer Threshold (RHT) using the HiRA-Memory Controller (MC). *AQUA* [62], similar to RRS, creates physical separation between aggressor and victim rows for efficient *Rowhammer* mitigation. In [75], authors challenge the effectiveness of RRS and propose the *Juggernaut* attack pattern. They introduce Secure Row-Swap (SRS), resilient to the *Juggernaut* attack, and scalable in Scale-SRS. SRS tracks accesses, identifies malicious patterns, and swaps aggressor rows to prevent the attack.

### 3.8 2023

Examining *Copy-on-Flip* [12] reveals the introduction of a software-based *Rowhammer* mitigation technique called Copy-on-Flip (CoF). CoF aims to strengthen ECC at the software level by isolating the vulnerable parts (potential victim rows) in memory and replacing them with different locations within the DRAM. Detection of these areas is achieved by monitoring the number of bits detected and corrected by ECC in that specific region. On the other hand, *PT-Guard* [61] offers a hardware-based mitigation technique to safeguard PTEs against *Rowhammer* attacks.

### 3.9 Defences in Categories

In concluding this section, we present the key categories that encompass prominent approaches to mitigating *Rowhammer*. These categories, which are not mutually exclusive, include: 1) Hardware-based, 2) Software-based, 3) Counter-based, 4) ECC-like, 5) Cell-focused, 6) Randomness-based, 7) Physical separation, 8) Restrictive measures, 9) Static analysis, and 10) Open Source Code.

**Hardware-based Defence Mechanisms.** Below, Table 12 illustrates the papers proposing hardware-based mitigation techniques to stop *Rowhammer*. Although, majority of them are pure hardware techniques, some are co-designed with software-based in order to leave a better effect to stop *Rowhammer*.

| | Paper(s) |
|---|---|
| 1. | Flipping Bits in Memory Without Accessing Them: An Experimental Study of DRAM Disturbance Errors [35] |
| 2. | Architectural support for mitigating row hammering in DRAM memories [30] |
| 3. | Flip Feng Shui: Hammering a needle in the software stack [57] |
| 4. | Counter-based tree structure for row hammering mitigation in DRAM [65] |
| 5. | DRAM row-hammer attack reduction using dummy cells [18] |
| 6. | Making DRAM stronger against row hammering [66] |
| 7. | TWiCe: Time window counter based row refresh to prevent row-hammering [41] |
| 8. | Reliably achieving and efficiently preventing Rowhammer attacks [26] |
| 9. | Rapid detection of Rowhammer attacks using dynamic skewed hash tree [72] |
| 10. | TWiCe: preventing row-hammering by exploiting time window counters [40] |
| 11. | An Effective DRAM Address Remapping for Mitigating Rowhammer Errors [34] |
| 12. | Detect DRAM Disturbance Error by Using Disturbance Bin Counters [74] |
| 13. | Graphene: Strong yet Lightweight Row Hammer Protection [50] |
| 14. | BlockHammer: Preventing RowHammer at Low Cost by Blacklisting Rapidly-Accessed DRAM Rows [79] |
| 15. | Stop! Hammer time: rethinking our approach to rowhammer mitigations [43] |
| 16. | Panopticon: A complete in-dram rowhammer mitigation [3] |
| 17. | Hammerfilter: Robust protection and low hardware overhead method for rowhammer [32] |
| 18. | Randomized row-swap: mitigating Row Hammer by breaking spatial correlation between aggressor and victim rows [59] |
| 19. | PROTRR: Principled yet Optimal In-DRAM Target Row Refresh [45] |
| 20. | CSI: Rowhammer-Cryptographic Security and Integrity against Rowhammer [29] |
| 21. | HiRA: Hidden Row Activation for Reducing Refresh Latency of Off-the-Shelf DRAM Chips [78] |
| 22. | Aqua: Scalable rowhammer mitigation by quarantining aggressor rows at runtime [62] |
| 23. | Scalable and Secure Row-Swap: Efficient and Safe Row Hammer Mitigation in Memory Systems [75] |
| 24. | Security Analysis of the Silver Bullet Technique for RowHammer Prevention [80] |
| 25. | PT-Guard: Integrity-Protected Page Tables to Defend Against Breakthrough Rowhammer Attacks [61] |

**Table 12.** Hardware-based Defence Mechanisms

**Software-based Defence Mechanisms.** Table 13 shows the list of works demonstrating mitigations for *Rowhammer* using software level techniques. Some of these defence mechanisms are co-designed with a hardware-based mitigations, however the majority of them are solely software-based.

| | Paper(s) |
|---|---|
| 1. | Flip Feng Shui: Hammering a needle in the software stack [57] |
| 2. | Drammer: Deterministic rowhammer attacks on mobile platforms [70] |
| 3. | ANVIL: Software-based protection against next-generation rowhammer attacks [2] |
| 4. | Can't touch this: Practical and generic software-only defenses against rowhammer attacks [6] |
| 5. | MASCAT: Stopping microarchitectural attacks before execution [25] |
| 6. | Secure Page Fusion with VUsion [49] |
| 7. | Reliably achieving and efficiently preventing Rowhammer attacks [26] |
| 8. | Throwhammer: Rowhammer attacks over the network and defenses [68] |
| 9. | GuardION: Practical mitigation of DMA-based rowhammer attacks on ARM [71] |
| 10. | Zebram: Comprehensive and compatible software protection against rowhammer attacks [38] |
| 11. | Stop! Hammer time: rethinking our approach to rowhammer mitigations [43] |
| 12. | CSI: Rowhammer-Cryptographic Security and Integrity against Rowhammer [29] |
| 13. | Aqua: Scalable rowhammer mitigation by quarantining aggressor rows at runtime [62] |
| 14. | Copy-on-Flip: Hardening ECC Memory Against Rowhammer Attacks [12] |

**Table 13.** Software-based Defence Mechanisms

**Counter-based Defence Mechanisms.** Several papers are based on keeping track of accesses using different types of counters and counting techniques. The list of papers in this category are illustrated in Table 14.

| | Paper(s) |
|---|---|
| 1. | Flipping Bits in Memory Without Accessing Them: An Experimental Study of DRAM Disturbance Errors [35] |
| 2. | Architectural support for mitigating row hammering in DRAM memories [30] |
| 3. | Flip Feng Shui: Hammering a needle in the software stack [57] |
| 4. | ANVIL: Software-based protection against next-generation rowhammer attacks [2] |
| 5. | Counter-based tree structure for row hammering mitigation in DRAM [65] |
| 6. | Making DRAM stronger against row hammering [66] |
| 7. | TWiCe: Time window counter based row refresh to prevent row-hammering [41] |
| 8. | Rapid detection of Rowhammer attacks using dynamic skewed hash tree [72] |
| 9. | TWiCe: preventing row-hammering by exploiting time window counters [40] |
| 10. | Detect DRAM Disturbance Error by Using Disturbance Bin Counters [74] |
| 11. | Graphene: Strong yet Lightweight Row Hammer Protection [50] |
| 12. | BlockHammer: Preventing RowHammer at Low Cost by Blacklisting Rapidly-Accessed DRAM Rows [79] |
| 13. | Stop! Hammer time: rethinking our approach to rowhammer mitigations [43] |
| 14. | Panopticon: A complete in-dram rowhammer mitigation [3] |
| 15. | Hammerfilter: Robust protection and low hardware overhead method for rowhammer [32] |
| 16. | Randomized row-swap: mitigating Row Hammer by breaking spatial correlation between aggressor and victim rows [59] |
| 17. | PROTRR: Principled yet Optimal In-DRAM Target Row Refresh [45] |
| 18. | Aqua: Scalable rowhammer mitigation by quarantining aggressor rows at runtime [62] |
| 19. | Scalable and Secure Row-Swap: Efficient and Safe Row Hammer Mitigation in Memory Systems [75] |
| 20. | Security Analysis of the Silver Bullet Technique for RowHammer Prevention [80] |

**Table 14.** Counter-based Defence Mechanisms

**ECC(-like) Defence Mechanisms.** Papers in this category employ ECC or ECC-like techniques to mitigate *Rowhammer* attacks. Table 15 presents the entries in this category.

| | Paper(s) |
|---|---|
| 1. | Flipping Bits in Memory Without Accessing Them: An Experimental Study of DRAM Disturbance Errors [35] |
| 2. | Rapid detection of Rowhammer attacks using dynamic skewed hash tree [72] |
| 3. | CSI: Rowhammer-Cryptographic Security and Integrity against Rowhammer [29] |
| 4. | Copy-on-Flip: Hardening ECC Memory Against Rowhammer Attacks [12] |
| 5. | PT-Guard: Integrity-Protected Page Tables to Defend Against Breakthrough Rowhammer Attacks [61] |

**Table 15.** ECC-like Defence Mechanisms

**Defence Mechanisms Focusing on Cells.** Table 16 demonstrates the list of papers that are either deactivating the vulnerable cells in DRAM or using *dummy* cells in efforts to prevent *Rowhammer* impacts on the victim machine.

| | Paper(s) |
|---|---|
| 1. | Flipping Bits in Memory Without Accessing Them: An Experimental Study of DRAM Disturbance Errors [35] |
| 2. | Can't touch this: Practical and generic software-only defenses against rowhammer attacks [6] |
| 3. | DRAM row-hammer attack reduction using dummy cells [18] |

**Table 16.** Defence Mechanisms Focusing on Cells

**Defence Mechanisms based on Randomness.** Probabilistic *Rowhammer* mitigations based on random choices fall under this category. We have considered [79] as it is based on using *bloom filters* which are probabilistic in nature. Similarly, [29, 61] are entered this category because of using MAC in the defence mechanisms. Table 17 demonstrates the list of papers in this category.

| | Paper(s) |
|---|---|
| 1. | Flipping Bits in Memory Without Accessing Them: An Experimental Study of DRAM Disturbance Errors [35] |
| 2. | Architectural support for mitigating row hammering in DRAM memories [30] |
| 3. | Flip Feng Shui: Hammering a needle in the software stack [57] |
| 4. | Making DRAM stronger against row hammering [66] |
| 5. | Reliably achieving and efficiently preventing Rowhammer attacks [26] |
| 6. | BlockHammer: Preventing RowHammer at Low Cost by Blacklisting Rapidly-Accessed DRAM Rows [79] |
| 7. | Hammerfilter: Robust protection and low hardware overhead method for r [32] |
| 8. | Randomized row-swap: mitigating Row Hammer by breaking spatial correlation between aggressor and victim rows [59] |
| 9. | CSI: Rowhammer-Cryptographic Security and Integrity against Rowhammer [29] |
| 10. | PT-Guard: Integrity-Protected Page Tables to Defend Against Breakthrough Rowhammer Attacks [61] |

**Table 17.** Defence Mechanisms based on Randomness

**Defence Mechanisms based on Physical Separation.** Table 18 lists the papers with mitigation techniques based on physical separation. Papers in this category are proposing ideas to mitigate *Rowhammer* by forcing a gap between potential attacker rows and potential victim rows.

| | Paper(s) |
|---|---|
| 1. | Drammer: Deterministic rowhammer attacks on mobile platforms [70] |
| 2. | Can't touch this: Practical and generic software-only defenses against rowhammer attacks [6] |
| 3. | Reliably achieving and efficiently preventing Rowhammer attacks [26] |
| 4. | Throwhammer: Rowhammer attacks over the network and defenses [68] |
| 5. | Zebram: Comprehensive and compatible software protection against rowhammer attacks [38] |
| 6. | Stop! Hammer time: rethinking our approach to rowhammer mitigations [43] |
| 7. | Randomized row-swap: mitigating Row Hammer by breaking spatial correlation between aggressor and victim rows [59] |
| 8. | Aqua: Scalable rowhammer mitigation by quarantining aggressor rows at runtime [62] |
| 9. | Scalable and Secure Row-Swap: Efficient and Safe Row Hammer Mitigation in Memory Systems [75] |

**Table 18.** Defence Mechanisms based on Physical Separation

**Defence Mechanisms based on Restriction.** This category contains mitigation techniques based on imposing and adding restrictions in the software level. Table 19 shows the list of papers in this category.

| | Paper(s) |
|---|---|
| 1. | Drammer: Deterministic rowhammer attacks on mobile platforms [70] |
| 2. | Protecting Page Tables from RowHammer Attacks using Monotonic Pointers in DRAM True-Cells [76] |

**Table 19.** Defence Mechanisms based on Restriction

**Defence Mechanisms based on Static Analysis.** This category contains the work using *Static Analysis* to detect *Rowhammer* and stop it. Table 20 lists the on only paper in the category.

| | Paper(s) |
|---|---|
| 1. | MASCAT: Stopping microarchitectural attacks before execution [25] |

**Table 20.** Defence Mechanisms based on Static Analysis

**Defence Mechanisms with Open Source Code.** Finally, we have a category for the papers that have made their code open source. Table 21 shows the list of the papers in this category.

| | Paper(s) |
|---|---|
| 1. | Throwhammer: Rowhammer attacks over the network and defenses [68] |
| 2. | GuardION: Practical mitigation of DMA-based rowhammer attacks on ARM [71] |
| 3. | Zebram: Comprehensive and compatible software protection against rowhammer attacks [38] |
| 4. | BlockHammer: Preventing RowHammer at Low Cost by Blacklisting Rapidly-Accessed DRAM Rows [79] |
| 5. | Randomized row-swap: mitigating Row Hammer by breaking spatial correlation between aggressor and victim rows [59] |
| 6. | CSI: Rowhammer-Cryptographic Security and Integrity against Rowhammer [29] |
| 7. | Aqua: Scalable rowhammer mitigation by quarantining aggressor rows at runtime [62] |
| 8. | Scalable and Secure Row-Swap: Efficient and Safe Row Hammer Mitigation in Memory Systems [75] |
| 9. | Copy-on-Flip: Hardening ECC Memory Against Rowhammer Attacks [12] |

**Table 21.** Defence Mechanisms with Open Source Code

# 4 Further Work

In this section, we will briefly explore works on *Rowhammer* that don't focus on attacks or defence mechanisms. These works approach the problem from different perspectives and conduct observations and experiments. The first work to mention is [63], which creatively utilizes *Rowhammer* to create a new Physically Unclonable Function (PUF) called *Rowhammer* PUF. *Rowhammer: A retrospective* [46] provides an overview of *Rowhammer* and various related works up to the publication date. Another review work, [73], examines *Rowhammer* from a hardware standpoint, delving into detailed hardware-level factors that influence *Rowhammer*'s impact. RowHammer-Aware Test (RHAT) [13] presents a novel framework for assessing DRAM's susceptibility to *Rowhammer* attacks, identifying vulnerable cells. The authors validate their findings by applying the framework to DRAMs from multiple manufacturers. An open-source tool called *mFIT*, showcased in [10], combines hardware-based and software-based techniques to measure *Rowhammer* vulnerability in DRAM.

The study presented in [15] focuses on simulating *Rowhammer* using the *gem5* simulator, enabling powerful and relative studies in a simulated environment. *Uncovering In-DRAM RowHammer Protection Mechanisms: A New Methodology, Custom RowHammer Patterns, and Implications* [23] is a comprehensive work on *Rowhammer* mitigations in DRAM modules, specifically TRR. The authors analyse 45 DRAM modules, introduce Uncovering TRR (U-TRR) as a reverse engineering approach for TRR, and verify its effectiveness by comparing attack patterns with previous research. In [60], the authors discuss the disadvantages of keeping DRAM internals obscure for customers, emphasising the need for transparency. And in [48], the authors introduce ALARM, a novel tool that utilises active learning to extract DRAM and *Rowhammer* behaviour characteristics. Another review paper, [47], categorises *Rowhammer* work into three main categories: exploiting *Rowhammer*, understanding (modelling) *Rowhammer*, and mitigating *Rowhammer*. *WhistleBlower* [24] conducts experiments on real-world FPGA configurations to evaluate hardware-based and software-based factors affecting *Rowhammer* attacks. They also introduce new factors. And *HammerSim* [19] is an open-source *Rowhammer* simulator in *gem5*.

# 5 Conclusion

In conclusion, *Rowhammer* attacks pose a significant security threat, exploiting the nature of hardware to manipulate computer memory and compromise both system integrity and confidentiality. Defence mechanisms have been proposed ever since the attack was introduced, but challenges remain in terms of false positives, resource overhead, and compatibility. Ongoing research and collaboration are necessary to develop robust countermeasures and safeguard against the evolving landscape of *Rowhammer* attacks.